\documentclass[authoryear]{elsarticle}
\usepackage{amssymb}
\usepackage{multirow}
\usepackage{booktabs}
\usepackage{fullpage}
\usepackage{lscape}
\usepackage{graphicx,psfrag,epsf} 
\usepackage{amsmath}
\usepackage{amsthm}
\usepackage{float}
\usepackage{rotating}
\usepackage{todonotes}
\usepackage{algpseudocode}
\usepackage{algorithm}
\usepackage{graphicx}
\usepackage{caption}
\usepackage{subcaption}

\usepackage{xcolor,colortbl}
\usepackage{url}
\usepackage{enumerate}
\usepackage{color}
\usepackage[shortlabels]{enumitem}
\usepackage{float}%

\DeclareMathOperator*{\argmin}{arg\,min}
\newcommand{\Expect}{{\rm I\kern-.3em E}}

\journal{International Journal of Production Economics}

\begin{document}

\begin{frontmatter}

\title{\LARGE Data-Driven Analytics for Benchmarking \\and Optimizing Retail Store Performance}

\cortext[mycorrespondingauthor]{Corresponding author: Ratna Babu Chinnam}

\author {Haidar Almohri}
\ead{haidar.almohri@wayne.edu}
\author {Ratna Babu Chinnam}%$^{*}$
\ead{ratna.chinnam@wayne.edu}
\address {Department of Industrial and Systems Engineering, Wayne State University, \\4815 Fourth Street, Detroit, MI 48202, USA}
\author {Mark Colosimo}
\ead{macolosimo@urbanscience.com}
%\ead{mark.colosimo@wayne.edu}
\address {Integrated Analytics, Urban Science, \\400 Renaissance Center, Detroit, MI 48243, USA}


\begin{abstract}
Growing competitiveness and increasing availability of data is generating tremendous interest in data-driven analytics across industries. In the retail sector, stores need targeted guidance to improve both the efficiency and effectiveness of individual stores based on their specific locations, demographics, and environment. We propose an effective data-driven framework for internal benchmarking that can lead to targeted guidance for individual stores. In particular, we propose an objective method for segmenting stores using a model-based clustering technique that accounts for similarity in store performance dynamics. The proposed method relies on an effective Finite Mixture of Regressions technique based on competitive learning for carrying out the model-based clustering with `must-link' constraints and modeling store performance. We also propose an optimization framework to derive tailored recommendations for individual stores within store clusters that jointly improves profitability for the store while also improving sales to satisfy franchiser requirements. We validate the methods using synthetic experiments as well as a real-world automotive dealership network study for a leading global automotive manufacturer. 
\end{abstract}


\end{frontmatter}

\section{Introduction}
\label{sec:intro}
Increasing global competition combined with product proliferation, dropping customer loyalties, and shrinking product life-cycles is changing the environment facing most companies today. None of the industries seem to be immune and retail and franchise sectors seem to be particularly hurting. The number of retailers filing for bankruptcy protection in the U.S. is headed toward its highest annual tally since the Great Recession in the 1920s \citep{bankrupcy}. While the reasons for bankruptcies and difficulties are several, many companies within these sectors lack comprehensive and effective performance management systems \citep{yu2009assessment}. These sectors have no doubt seen tremendous efficiencies from employing in-store technologies (e.g., scanning systems, enhanced point-of-sale systems, self-service lines) and information technology to drive upstream operations (e.g., warehousing, logistics, and manufacturing) \citep{king2004modeling}. 
%The recent 2017 RIS/Gartner Retail Technology Study reveals focus by retailers on unified commerce, personalized marketing, and customer engagement \citep{bus1}. 
While these technologies and the associated Retail Information Systems (RIS) are effective in helping to manage core store activities (like inventory control and logistics), these chains also need analytics platforms for performance management \citep{rigby2011future, nash2013customer}.

With the increasing availability of data and technologies to store and process the data, business analytics has been experiencing growing attention among the researchers and is  transforming the way businesses operate in many sectors \citep{sun2017business, MGI2016}. In the context of retail and franchise networks, stores need customized analytical guidance to improve profitability and sales of individual stores based on their specific location, demographics, and environment. Executives are unanimous in voicing their concerns over the lack of methods to assess store specific issues and derive equally specific insights \citep{bucklin1999commercial}. Such guidance is key to important strategic management decisions, including evaluation, promotion and development. Strategic resource-allocation decisions, such as advertising budgets, store expansions/closings, are also based on management's understanding of store performance drivers \citep{thomas1998process}. 

Parsons makes the distinction between store efficiency and effectiveness. Efficiency refers to the relationship between inputs and outputs, while effectiveness focuses on outputs relative to a particular objective \citep{parsons1994productivity} . Productivity is the combination of efficiency and effectiveness and is the focus of this study.  Thomas et al. note that productivity studies demand several careful considerations \citep{thomas1998process} . First, relevant individual store differences must be considered within the platform to take into account advantages and disadvantages of particular stores, e.g., location, competitive intensity \citep{kamakura1996productivity}. Second, development is much more effective when specific practices can be observed and transferred to other stores. Effective practices should be identified, described, and used as benchmarks for less efficient stores. This is a key focus of this study and the proposed method attempts to attribute (through modeling) variation in performance across a set of stores to the way the stores are managed. In particular, we will demonstrate how the key performance indicators (KPIs) are managed using resources. Third, a distinction must be made between resources under the control of store vs. those they have little or no influence over (e.g., local land/rent costs). In our proposed approach, we emphasis deriving recommendations around ``actionable" KPIs (e.g., increasing advertising budget) vs. not so actionable KPIs (e.g., changing store location or reducing overhead). Fourth, more than one outcome usually needs to be considered because stores are responsible for multiple and sometimes conflicting performance measures (e.g., dealership sales might be more important to an automotive manufacturer vs. profits to the owner's of the dealership). This is also addressed in this study through the use of multi-objective optimization methods for deriving recommendations. 

Numerous methods have been proposed for evaluating retail efficiency of individual stores  \citep{balakrishnan1994efficiency, kamakura1996productivity}. Store performance could be influenced by trade area demographic factors \citep{ingene1980market}, level of competition \citep{ghosh1984}, retail atmospherics \citep{jain1979evaluating}, and promotions \citep{walters1988structural}. The effect of internal retail environment including level of service and extended store hours as well as overlapping trading areas have been studied\citep{pauler2009assessing,kumar2000effect}. As for methods, Data Envelopment Analysis (DEA) has been used extensively for benchmarking performance of retail stores \citep{kamakura1996productivity,donthu1998retail,vyt2008retail}. DEA is a nonparametric method in operations research and economics to empirically measure productive efficiency of decision making units. The main advantage of DEA is its ability to accommodate a multiplicity of inputs and outputs and the absence of a need to explicitly specify a mathematical form for the production function. DEA develops a function whose form is determined by the most efficient producers and differs from the Ordinary Least Squares (OLS) statistical regression technique that bases comparisons relative to an average producer. A major drawback of DEA however is that the model specification and inclusion/exclusion of variables can affect the results. It allows individual units to employ nearly arbitrary weights/importance to inputs and performance metrics to showcase their relative efficiency. For these reasons, it is more applicable for activities such as outlet termination \citep{vyt2017towards}. Instead, we take the statistical regression approach for deriving improved and actionable recommendations for retail performance management. %Another common approach that is adopted by economists for evaluating the efficiency of retail stores is using translog (Transcendental Logarithmic) cost function, a second-order approximation to a cost function that can be used to model how a firm combines inputs to produce outputs \citep{caves1982economic, kamakura1996productivity}. 

In summary, the focus of this manuscript is to facilitate improvement in the performance of individual stores by relying on a data-driven approach to internal benchmarking. In particular, the goal is to identify factors driving automotive dealership performance in comparison with ``similar" dealerships and relying on optimization to derive tailored recommendations. The problem was brought to our attention by a global leader in providing automotive dealership location and network analysis to many automotive original equipment manufacturers (OEMs). Our contributions are as follows: 1) We propose an objective method for segmenting stores using a model-based clustering technique that accounts for similarity in store performance dynamics -- this is being done to better account for the considerations promoted by \citep{thomas1998process}, 2) We propose an effective Finite Mixture of Regressions (FMR) technique based on competitive learning for carrying out the model-based clustering and modeling store performance called Mixture Models with Competitive Learning (MMCL), and 3) We propose an optimization framework to derive tailored recommendations for individual stores within store clusters that jointly improves profitability for the store while also improving sales to satisfy OEM/franchiser requirements. We illustrate the methods using synthetic experiments and a real-world dataset from a leading global OEM. 

The rest of this manuscript is organized as follows: Section \ref{sec:adp} provides an overview of automotive dealership performance management and the proposed overall approach for performance management. Section \ref{sec:fmm} provides background information regarding finite mixture models. Section \ref{sec:meth} describes the proposed mixture model with competitive learning (MMCL) for the problem of finite mixture of regressions (FMR) under group structure constraints. Section \ref{sec:Verify} presents results from synthetic experiments to validate the effectiveness of \textit{MMCL}. Section \ref{sec:MOO} illustrates our method for deriving tailored recommendations using cluster specific component models from \textit{MMCL} using multi-objective optimization. Section \ref{sec:Case} describes results from a dealership case study. Finally, section \ref{sec:conc} offers some concluding remarks and directions for future research.

\section{Automotive Dealership Performance Management}
\label{sec:adp}

In the automotive industry, dealer efficiency and effectiveness are key factors for obtaining and maintaining competitiveness for OEMs. This is just as critical for the well-being and durability of dealers, for most dealerships tend to be franchises (in the U.S. and much of the world) that have a contract with an automotive OEM that allows them to sell its products. It is critically important to establish analytics platforms for assessing the productivity of the dealer network that not only is  useful for the OEM but also provides customized guidance to individual dealerships. Automotive OEMs usually assess dealership performance according to market share and plan incentive systems by assigning annual sales targets to each dealership. However, performance assessment based on a simplistic comparison between the dealership and national or state average market share can not only lead to ineffective sales targets but could also compromise the productivity of the dealer and the competitiveness of the OEM  
\citep{biondi2013new}. As noted by Biondi and co-authors, this kind of assessment does not take into account either the availability or the utilization of resources. Dealership \texttt{A} may be more efficient than dealership \texttt{B} according to the market share method, although \texttt{A} can obtain a higher output than \texttt{B} (i.e., sell more vehicles) merely because \texttt{A} has a more consolidated presence in the territory (i.e., has had a sales mandate for longer) and/or is located in a more favorable geographical market (e.g., where the brand enjoys more loyalty). Therefore, a more objective modeling and analysis methodology is necessary for evaluating and improving the performance of dealerships.

%\begin{figure}[H]
%	\centering
%	\includegraphics[trim={2.5cm 19.75cm 2.5cm 5.2cm},clip,scale= 0.9]{FlowChart.pdf}
%	\caption{\footnotesize{Retail Performance Analytics Platform}}
%	\label{figure:1}
%\end{figure}%

The company wanted to develop a dealership performance management analytics platform to analyze monthly operations and financial data (including information on sales staffing levels/tenure, product assortment/mix, dealer services (e.g., financing, trade-ins, collision repair), advertising budgets/mix, service bays/technicians etc.) from thousands of dealers in the U.S. to understand factors that can jointly improve profitability for the dealership while also improving vehicle sales to satisfy OEM requirements. In the absence of objective data-driven analytics platforms, dealerships mostly rely on experienced consultants and ad hoc guidance from field personnel. We propose a model-based ``competitive learning" method for clustering the stores into similar groups for benchmarking. The proposed method is generic and can be utilized in combination with a variety of both internal (e.g. inventory, advertising) and external data sources (e.g., demographics and local competition). 

\begin{figure}[t]
	\centering
	\includegraphics[scale= 0.55]{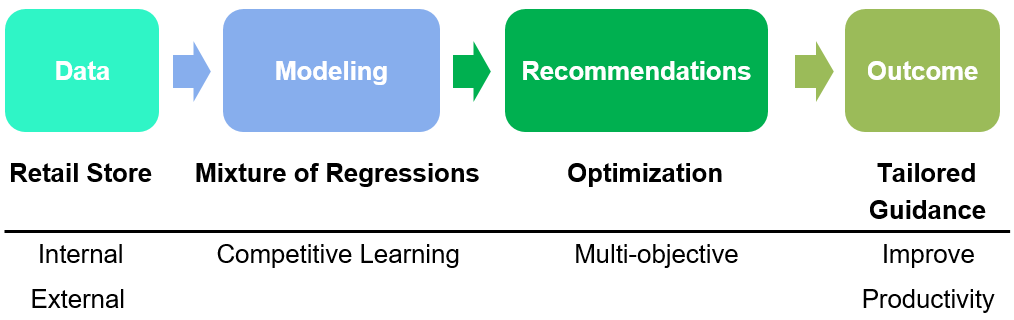}
	\caption{\footnotesize{Retail Performance Management Analytics Process}}
	\label{figure:1}
\end{figure}

Figure \ref{figure:1} illustrates our overall approach to retail performance management analytics. The process starts with acquiring appropriate data from both internal and external sources. Next, the stores are clustered by our proposed MMCL algorithm and the important KPIs that have the highest influence on the key process objectives (KPOs) (e.g. dealership profitability and sales effectiveness) are identified. The resulting relationships are employed by the proposed multi-objective optimization (MOO) method to determine the optimal settings for each KPI to achieving the desired objectives. The outcome of the optimization and MMCL phases provides tailored guidance to stores for improving their performance and productivity. Details of each phase are discussed in more detail within later sections of the manuscript. 

\section{Finite Mixture Models (FMM)}
\label{sec:fmm}

This section provides an overview of finite mixture models for the purpose of segmenting/clustering stores using a model-based clustering technique that accounts for similarity in store performance dynamics. The objective is to cluster the stores into a number of homogeneous groups for benchmarking and deriving more effective recommendations. For example, it might be inappropriate to benchmark a rural dealer with an urban dealer in a large metropolitan city. This is above and beyond the normal practice of examining stores based on regional location. For example, it is a common practice in the U.S. for the automotive OEMs to look at the continental U.S. as several major regions (e.g., North-East, Midwest, South-East, South-West, and West) due to significant differences in weather and other purchasing patterns. While we too recommend regional analysis, there is still room for improving the performance modeling by further clustering the regional stores into a number of smaller homogeneous store groups.

In mixture models, ``components" are introduced into the mixture model to allow for greater flexibility in modeling a heterogeneous population that is unable to be modeled by a single model. The hope is that this form of clustering would allow for more effective modeling and comparison of stores within individual clusters and for recognition/extraction of effective practices to be used as benchmarks for less efficient stores.  

\subsection{Mixture Models for Clustering}
Clustering is the process of finding subsets of a dataset based on ``similarity”, where members of the subsets are similar and members across subsets are dissimilar \citep{guha2016clustering}. There are several algorithms that have been proposed for the clustering problem \citep{xu2005survey}. Traditional clustering methods for the most part are heuristic techniques derived from empirical methods and have difficulty taking into account the characteristics of clusters (shapes, proportions etc.). Finite mixture models have attracted much attention in recent years for clustering.  McLachlan and Basford were the first to highlight the usefulness of mixture models as a way of providing an effective clustering of various datasets under a variety of experimental designs \citep{mclachlan1988mixture}. FMM offer considerable flexibility and permit certain classical criteria for vigorous analysis and have been widely used for market segmentation and similar studies \citep{andrews2011comparison, wedel2002market, sarstedt2008market, tuma2013finite}.

% Reviwers generally care about applications in the same domain. Given that we have a lot of references from this domain, no need to list application areas such as image processing or gene screening.

In mixture models, it is assumed that observations are generated according to several probability distributions (a.k.a. components) with certain parameters. Data points in each distribution are assumed to form a cluster. The general framework of an FMM is of the following form:
\begin{align}
f(x) =  \sum_{k=1}^{K} \alpha_k f(x_k;\theta_k) 
\end{align}
where $k$ is the number of mixture components, $\alpha_k$ is the mixing weights i.e. the proportion of component $k$ ($\alpha_k >0$ and $\sum_k{\alpha_k} = 1$), $\theta_k$ is the set of parameters for the $k$th component, and $f(x_k;\theta_k)$ is the distribution of the $k$th component. Each componentm, $k$, is assumed to come from a unique probability distribution $f(x_k;\theta_k)$, with probability $\alpha_k$ that an observation comes from component $k$. In the case of mixture of Gaussian distributions,  $f(x_k;\theta_k)\sim \mathcal{N}(\mu,\Sigma)$, where $\mu$ and $\Sigma$ denote the mean and covariance matrix for each component distribution, respectively.

\subsection{Finite Mixture of Regressions (FMR)}
Given our desire to model the performance of dealerships as a function of the KPIs, we need to explicitly handle the presence of dependent variables (e.g., standardized dealership sales and profits calculated relative to averages).  Quadnt and Ramsey proposed a method of moment algorithm to estimate the parameters of FMM with the presence of a dependent variable \citep{quandt1978estimating}. In this setup, it is assumed that the data points ($X \in{\rm I\!R}^n$) have an associated dependent variable ($Y \in{\rm I\!R}$), and the relation between $X$ and $Y$ is linear: $X=\beta Y$. The finite mixture of regression (FMR) models have been studied and used in various fields and applications 
\citep{bar1978tracking, andrews2003retention, bierbrauer2004modeling}, particular in the form of a mixture of ``linear" regression models mainly because of advantages such as simplicity and interpretability. It is assumed that the response variable, $y$, can be predicted using a finite $(K)$ number of linear regression models:
\begin{align}
y_i = \sum_{k=1}^{K} \alpha_k \phi \big(\frac{x_i\beta_k-y_i}{2\sigma_k^2}) \qquad  k =1,...,K
\end{align}

There are a number of methods to estimate the parameters of FMR models. Maximum likelihood, method of moments, Gibbs sampling, and Bayesian approach are some examples \citep{quandt1978estimating, viele2002modeling, faria2010fitting, hurn2003estimating}, each with their own advantages and disadvantages \citep{melnykov2010}. 

\subsection{Mixture Models with Constraints}
\label{sec:mmconst}
In some applications, however, instead of individual observations, ``groups" of observations need to be clustered together. In other words, it may be desirable to force a group of observations to stay in the same cluster. Given the current context of clustering dealerships, we need all the data from a particular dealer to be clustered within the same model and not distributed across different models. Let $D$ be the set of data points, and $M$ be the number of all distinct groups within $D$. Define $S$ as the set that holds all the group members: $S=\{s_i\}_{i=1}^{M}$. Each group $s_i$, with $n_i$ observations, has to retain its members when assigned to a cluster. The goal is to assign each $s_i$ to one of $K$ clusters. This problem is similar to what is known as ``clustering with must-link constraint", which is introduced by Wagstaff and Cardie (2001) in the literature \citep{Wagstaff:2001}. We refer to \citep{davidson2007survey} and \citep{basu2009constrained} for further details and applications of constrained clustering.

To the best of our knowledge, all the existing algorithms have solved the problem of clustering with must-link constraint in unsupervised/semi-supervised settings, meaning that the observations lack (or partially lack) the dependent variable. We could not find any work that addresses FMR with must-link constraints. Hence, we propose a non-parametric ``competitive learning" approach for assigning predefined groups/blocks of observations to clusters during model-based regression. This method is outlined in the next section.
\section{Mixture Model with Competitive Learning (MMCL)}
\label{sec:meth}
In this section, we propose an algorithm that relies on ``competitive learning" for FMR with group structure constraints, labeled Mixture Model with Competitive Learning (MMCL). %It is a distance minimization algorithm that minimizes a similarity measure to pick the best model for each observation group during regression. 
Later sections illustrate how this algorithm can be employed to address the problem of automotive dealership clustering and performance management. We also offer guidance on parameter initialization for \textit{MMCL}. 

\subsection{Competitive Learning}
Competitive learning is a form of unsupervised learning originating in the domain of artificial neural networks, in which nodes of the network compete for the right to respond to a subset of the input data \citep{rumelhart1988parallel}. %A variant of Hebbian learning, standard competitive learning algorithms work by increasing the specialization of each node in the network (typically composed of a single layer of neurons) and is well suited to finding clusters within data.
There are three basic elements to the standard competitive learning rule \citep{haykin2009neural}: 1) A set of neurons that are all the same except for some randomly distributed synaptic weights, and therefore respond differently to a given set of input patterns; 2) A limit imposed on the `strength' of each neuron, and 3) A mechanism that permits neurons to compete for the right to respond to a given subset of inputs, such that only one output neuron is active at a time. 
%Typically, the neuron that wins the competition is called a ``winner-take-all" neuron. Accordingly, during the training cycle, the individual neurons of the network learn to specialize on ensembles of similar patterns and in so doing become ``feature detectors" for different classes/clusters of input patterns.

In what follows, we adapt the standard competitive learning algorithm to the more general model setting of FMR (with grouping structure). 

\subsection{MMCL Algorithm}
As noted in section \ref{sec:mmconst}, let $D$ be the complete set of data points and $M$ the number of distinct groups within $D$ (e.g., data from each dealer forms an observation group). Define $S$ as a set that holds all the groups: $S=\{s_i\}_{i=1}^{M}$. Each group $s_i$ (that has $n_i$ observations in it), has to retain all its members when assigned to a cluster, forming the group structure constraints. If the desired number of clusters is $K$, the goal is to assign each $s_i$ to one of $K$ clusters. Ideally, the dataset is partitioned into training and testing datasets, where a subset of the data from each group is stored in the testing dataset for testing and improving model robustness. 

The Psuedo code for \textit{MMCL} is provided in Algorithm \#1. %%vspace{4mm}

\begin{algorithm}
	\caption{MMCL for FMR with Group Structure Constraints}
	\label{CLalgorithm}
	\begin{algorithmic}[T]
		\Procedure{\textendash $\>$Competitive Learning}{}\\
		\textbf{Input} the dataset $D$ \\
		\textbf{Initialize} $\>AIC$ to a large value (e.g., $10^{10}$) and $\epsilon$ to a small value (e.g., 0.001)
		\State Randomly select $K$ groups ($s_i$) for initializing each of the cluster component models
		\Repeat  
		\State $AIC_{Old} = AIC $
		\State Learn $K$ component models using observations assigned to $K$ clusters
		\For{each group $s_i \in S$}
		\State Make predictions for selected group using each of $K$ models and record AIC
		\State Assign $s_i$ to cluster with the least AIC
		\EndFor
		\State Calculate $Cluster.AIC_{j}$, $\forall{j={1,\dots,K}}$
		\State $AIC =\sum_{j=1}^K{Cluster.AIC_{j}}$
		\Until{convergence} (i.e., {$ | \frac {AIC - AIC_{Old}} {AIC_{Old}} | < \epsilon $ })\\
		\textbf{Output} Cluster assignment for each group $s_i$ 
		\EndProcedure
	\end{algorithmic}
\end{algorithm}

Assuming that the number of components, $K$, is known, the proposed \textit{MMCL} algorithm starts by randomly selecting $K$ groups (out of $M$) for initializing the clusters, and fitting one model using  observations in each group to get a function $f_i(x;\theta)$ for each $i \in {1,...,K}$. In the event the individual groups are too small to learn the initial model for each cluster component, one can randomly assign multiple groups to each component for initialization. This approach is generic in a sense that the component model can be of any type such as linear regression, LASSO, multi-layer perceptron, support vector machine, etc. Next, all the groups $s_i \in S$ are selected one at a time, and are predicted using each of the $K$ models. The group is assigned to one of the $K$ cluster components that produces the best performance; this is ``competition" portion of the \textit{MMCL}. 

For performance assessment, one can employ several criteria based on the component model, such as SSE of prediction, Bayesian Information Criteria (BIC) \citep{schwarz1978estimating}, or Akaike Information Criteria (AIC) \citep{akaike1998information}, on a held out testing dataset. Performance is evaluated on the testing dataset and not the training dataset to reduce the chance of over fitting. In our experiments with linear regression component models, AIC provided robust performance. Founded on information theory, given a collection of models for the data, AIC estimates the quality of each model relative to each of the other models \citep{akaike1998information}. In doing so, it deals with the trade-off between the goodness of fit of the model and the complexity of the model and is one of the most common model selection procedures available in most statistical software packages \citep{chaurasia2012using}. Akaike stated that
modeling is not only about finding a model which describes the behavior of the observed
data, but its main aim is predicated as a possible good, and the future of the process
is under investigation \citep{infocr2014}. The AIC is calculated as:
\begin{align}
AIC = -2l_f(\hat {\theta})+2k
\end{align}
where $l_f(\hat {\theta})$ is the maximum value of the likelihood function of the model with parameters $\hat {\theta}$. Given that models with minimum AIC are preferred, AIC employs the term $2k$ to penalize complex models with more parameters.

Without loss of generality, in the rest of the manuscript, we assume that AIC is the criterion for cluster component competition within \textit{MMCL}. 

After assigning each group to the cluster with the best (i.e., minimum) AIC, the ``Overall AIC" is calculated as the sum of the Cluster AICs: $Overall \, AIC = \sum_{j=1}^K{Cluster.AIC_{j}}$. In the next iteration, the process is repeated and this time the $K$ component models are learnt using all the aggregated observations (that consist of several groups) within each of $K$ clusters; this is the ``learning" portion of \textit{MMCL}. The updated models will compete in the same fashion as described above, to select the cluster members that produce the lowest AIC. This process is repeated until algorithm convergence (e.g., no changes in cluster group memberships between iteration $t$ and $t+1$ and/or the resulting Overall AIC) or if the maximum number of iterations has been reached. This is a common approach to stop searching in most of the meta-heuristic optimization methods \citep{safe2004stopping}.

Let us denote $\mathcal{M}_j$ as the $j^{th}$ model (that is built using the observations in $j^{th}$ cluster) and $\mathcal{A}_{\mathcal{M}_j,s_i}$ as the AIC value resulting from predicting the response values in group $s_i$ using model $\mathcal{M}_j$. The ultimate goal is to predict a label $c^{(i)}$ for each group $s_i$. We can write the algorithm in the form of an optimization problem as follows:
\begin{align} \label{eq:1}
&c^{(i)}=\argmin_j{\mathcal{A}_{\mathcal{M}_j,s_i}}
\end{align}
It can be seen that \textit{MMCL} is a special version of the famous \textit{k-means} clustering \citep{steinhaus1956division}, with the distance defined as the AIC of predictive models.

In the case of a linear regression component model, Eqn. \ref{eq:1} can be written as:
\begin{align} \label{eq2}
&c^{(i)}=\argmin_j{nlog\left|\frac{RSS_{\mathcal{M}_j,s_i}}{n_i}\right|+2p}
\end{align}
where $p$ is the number of parameters in the model and $RSS_{\mathcal{M}_j,s_i}=\sum_{l=1}^{n_i}{(\hat{y}_{\mathcal{M}_j,s_{il}}-y_{s_{il}})^2}$ is the  residual sum of squares resulting from predicting response variables in group $s_i$ using model $\mathcal{M}_j$. It is easy to observe that the with a minor modification, any stopping criteria and model selection technique can be used. 

\subsection{Initializing MMCL Parameters}
There are three main parameters that need to be selected prior to applying \textit{MMCL}: 
\begin{enumerate}[1),nolistsep]
\item Number of clusters $K$: An existing method such as BIC, Calinsky criterion, Gap Statistics, etc. can be used. With the existence of response variables (supervised learning), it is recommended to divide the data to train and test sets and select $K$ that yields the best result. 
\item Initializing $K$ clusters:
This is a critical step as it effects both the convergence and effectiveness of the algorithm. 
One of the existing methods such as \textit{k-means++} \citep{arthur2007k} can be adopted to develop an algorithm that wisely selects the initial cluster groups. 
%\textit{k-means++} has two steps for selecting the cluster centroids:
%\begin{enumerate}[nolistsep,label=\alph*.]
%\item Select one center from the data points uniformly at random
%\item Compute the distance $\mathcal{D}(x)$ between each data point and the centers that have already been selected
%\item Select a new center with probability proportional to $\mathcal{D}(x)^2$
%\item Repeat steps (a) and (b) until all $K$ centers are selected
%\item Apply \textit{k-means} using the selected points as initial cluster centroids
%\end{enumerate}  

Inspired by \textit{k-means++}, the following initialization algorithm (labeled \textit{MMCL++}) is developed to  select the initial groups. It tries to smartly choose the groups so that the selected groups have maximum dissimilarity. This is achieved by selecting a group $s_j \in{S}$ at random and predicting all the remaining groups using the model $\mathcal{M}_j$ that is developed by the observations in that group. The quality of prediction for all the remaining groups using $\mathcal{M}_j$ is evaluated, and the group $s_i, i\neq j$ that $\mathcal{M}_j$ has the least power predicting it is identified as the candidate that has the maximum distance with $s_i$. Again, different criteria such as correlation between $s_i$ and $s_j$, RSS of $\mathcal{M}_j,s_i$, etc. can be used for this purpose. 

\begin{algorithm}
\caption{\textit{MMCL++ Initialization Algorithm}}
\begin{algorithmic}
\Repeat 
\State Select one observation group $s_j \in{S}$ at random
\State Learn model $\mathcal{M}_j$ using the observations in $s_j$
\State Predict remaining groups $s_i, i\neq j$ using  $\mathcal{M}_j$ and calculate $\mathcal{A}_{\mathcal{M}_j,s_i}, \forall\,{i=\left\{1,\dots,M\right\}} \, \wedge i\neq j$
\State Select a new group $s_i$ that has $max(\mathcal{A}_{\mathcal{M}_j,s_i}), i=\left\{1,\dots,M\right\} \, \wedge i\neq j$
\Until {$K$ groups are selected}
\end{algorithmic}
\end{algorithm}

\item Parameters of the models $\theta$: With the current version of the algorithm, the parameters of the models can be only optimized using cross validation.
   
\end{enumerate}

\section{MMCL Validation: Synthetic Experiments}
\label{sec:Verify}

To evaluate the effectiveness of the proposed \textit{MMCL} algorithm for FMR with group structure constraints, we employ Monte Carlo simulation experiments.
 
\subsection{Experiment Setup}
For the synthetic experiments, for ease of exposition, we use linear regression as the modeling technique and investigate the different parameters on the result is investigated. The experiment is conducted for the case $K=2$ (number of clusters). Covariates ($X$) for each cluster are generated by drawing samples from a bivariate Gaussian distribution: $X \sim \mathcal{N}(\boldsymbol{{\mu}} , \boldsymbol{{\Sigma}})$, with zero mean and a diagonal covariance matrix with unit variance.
%$\boldsymbol{\mu}= \begin{pmatrix}0\\0\end{pmatrix} $, and $\boldsymbol{\Sigma}= \begin{pmatrix}\mathcal{N}(0,1)&0\\0&\mathcal{N}(0,1)\end{pmatrix} $. 

\begin{table}[H]
	\centering 
	\caption{Monte Carlo Simulation Parameters}
	\label{table:1}
	\begin{tabular}{l|c|c|c|c|}
		\cline{2-5}
		& \multicolumn{1}{c|}{$N$} & $S$  & Noise Level ($\epsilon$)& $d^2$ \\ \hline
		\multicolumn{1}{|l|}{Cluster 1} & 300 & 5 & \multirow{2}{*}{{(}0.5, 1, 2, 4, 6{)}} & \multirow{2}{*}{{(}0.2, 0.6, 1.8{)}} \\ \cline{1-3}
		\multicolumn{1}{|l|}{Cluster 2} & 300 & 15 & & \\ \hline
	\end{tabular}
\end{table}

Referring to the Monte Carlo simulation parameters outlined in Table \ref{table:1}, $N=300$ is the total number of observations in each cluster and $S$ is the number of groups (blocks) in each cluster. Essentially, there will be 60 observations per group in cluster 1 and 20 observations per group in cluster 2. The response variable for each observation is generated by: $y_i= {X_i}^{'}\beta+Noise\; level (\epsilon)$. $\epsilon$ parameter is used to control the amount of noise (uncertainty) added to the response variable $y$. It can be also seen as a parameter that controls the signal to noise ratio (SNR).

To study the effect of the degree of similarity between $\beta$s, the Euclidean distance ($d^2$) between ${\beta_1}$ and ${\beta_2}$ is calculated to control the level of separation for the two clusters (in the response domain). 
\begin{align}
&d^2 =  \left| \left| \beta _{1}-\beta _{2}\right| \right| ^{2}=\left| \left| \beta _{1} \right| \right|^{2}+\left| \left| \beta _{2} \right| \right|^{2}-2\left\langle \beta _{1},\beta _{2}\right\rangle = 2(1-{r_{12}})
\end{align}
where $r_{12} = \left\langle \beta _{1},\beta _{2}\right\rangle$, assuming that  $\left| \left| \beta _{1} \right| \right|^{2}=\left| \left| \beta _{2} \right| \right|^{2}=1$ ($\beta$s have $l_2$ norm of one). If $R$=[1 $r_{12}$; $r_{12}$ 1], and the matrix $B$ is the Cholesky decomposition of $R$, then the $i$th row of $B$ is $\beta_i$, with square distance $d^2$ between $\beta_1$ and $\beta_2$. Obviously, the smaller $d^2$, the closer the $\beta$s, and it is harder to separate the clusters.

\subsection{Simulation Results}
\begin{figure} %[ht]
\begin{subfigure}{.5\textwidth}
\centering
\includegraphics[width=0.9\linewidth]{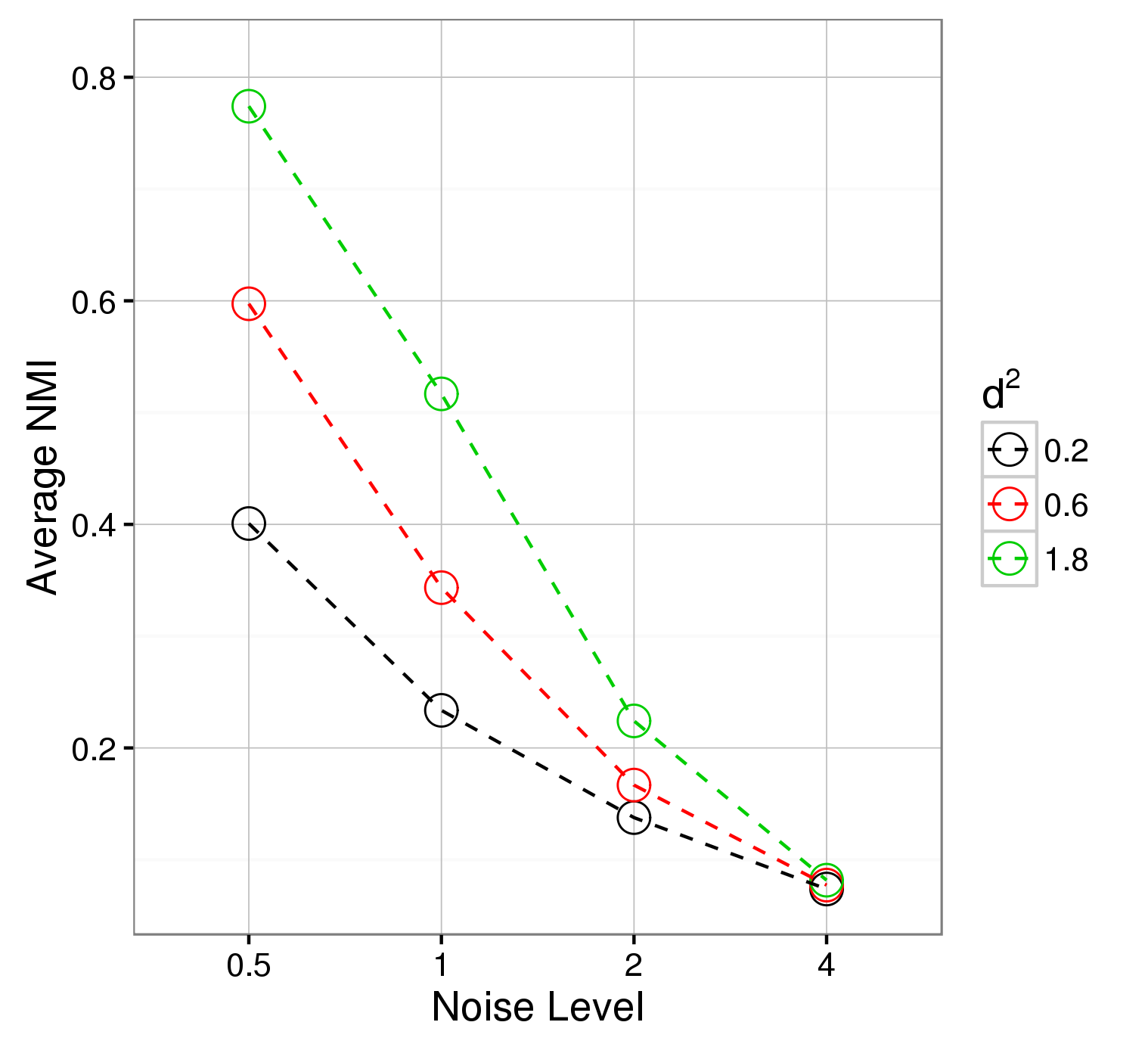}
\caption{\footnotesize{}}
\label{figure:sfig1}
\end{subfigure}%
\begin{subfigure}{.5\textwidth}
\centering
\includegraphics[width=0.9\linewidth]{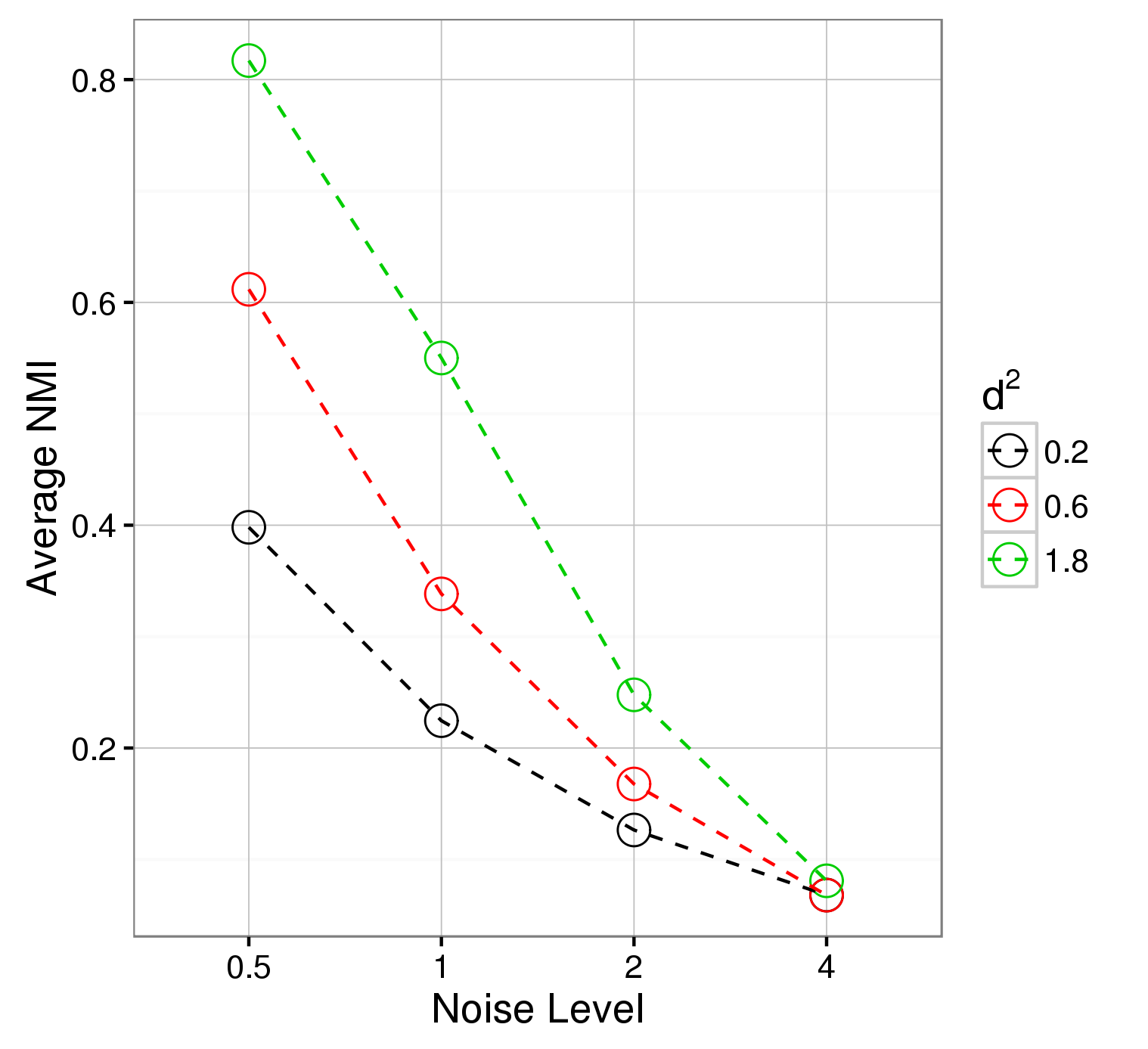}

\caption{\footnotesize{}}
\label{figure:sfig2}
\end{subfigure}
\begin{subfigure}{.5\textwidth}
\centering
\includegraphics[width=0.9\linewidth]{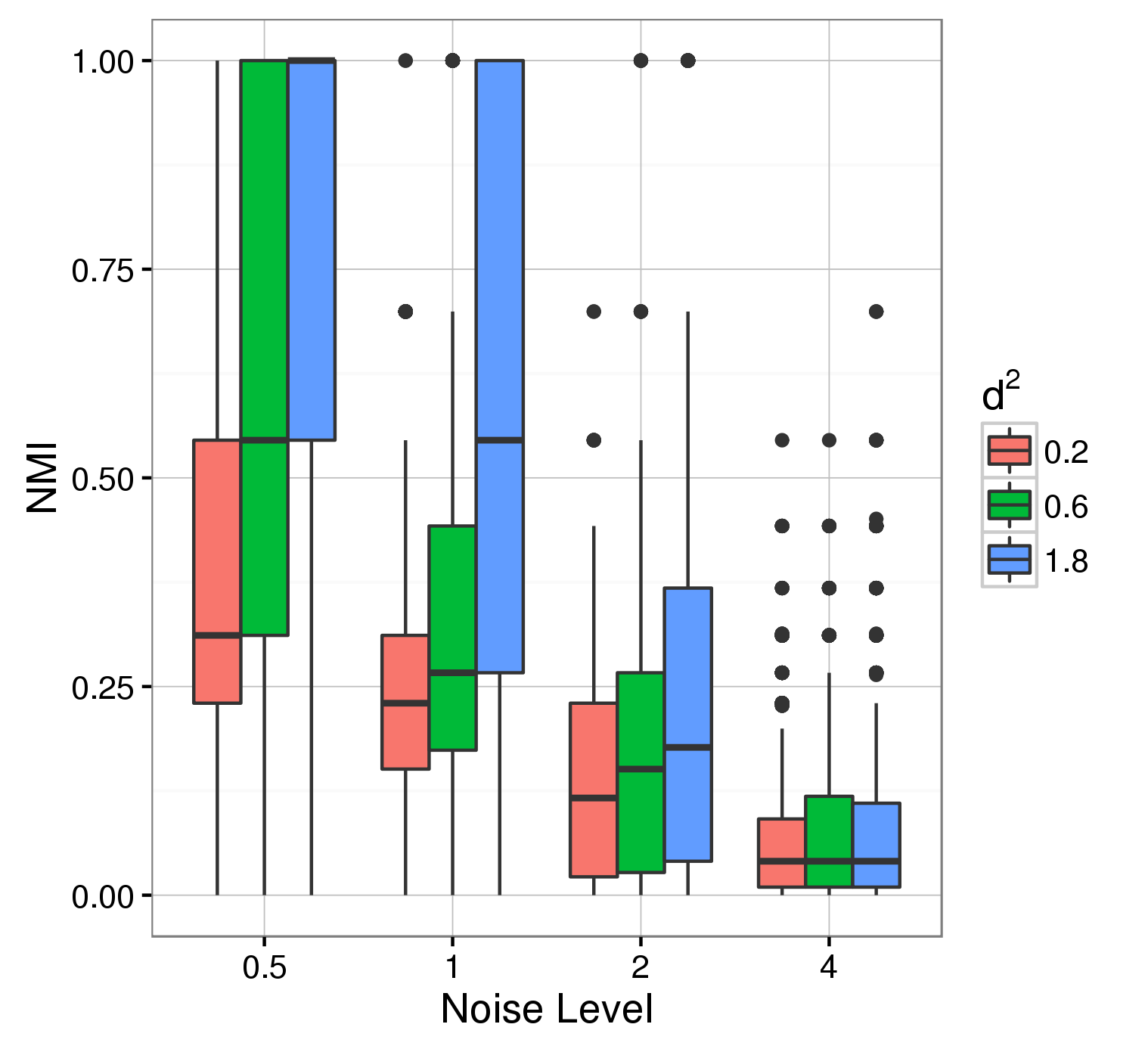}

\caption{\footnotesize{}}
\label{figure:sfig3}
\end{subfigure}
\begin{subfigure}{.5\textwidth}
\centering
\includegraphics[width=0.9\linewidth]{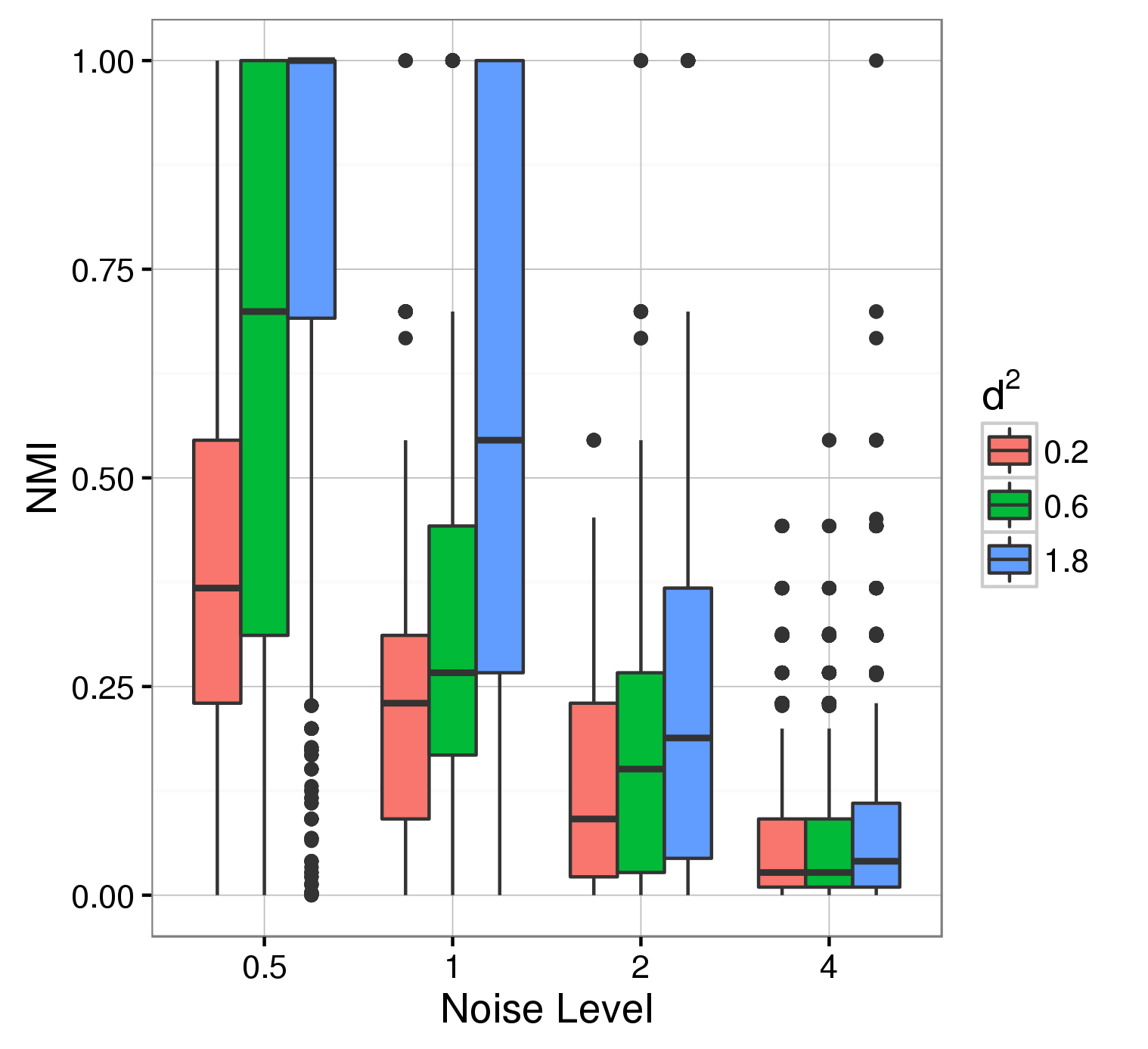}

\caption{\footnotesize{}}
\label{figure:sfig4}
\end{subfigure}
\caption{MMCL Performance on Synthetic Datasets: (a) Average NMI for different noise levels and $d^2$ (random initial groups assignment). (b) Average NMI for different noise levels and $d^2$ using \textit{MMCL++}. (c) Distribution of NMI for different noise levels and $d^2$ (random initial groups assignment). (d) Distribution of NMI for different noise levels and $d^2$ using \textit{MMCL++}.}
\label{figure:MMCL_SyntehticResult}
\end{figure}

The Monte Carlo simulations are repeated 1000 times for each pair of $d^2$ and $\epsilon$. Normalized Mutual Information (NMI) is used for assessing the clustering accuracy. NMI is a widely used technique in evaluating the clustering result when the true labels are available. The advantage of using NMI is that it is independent of permutation, meaning that the label switching does not affect the NMI score. It is bounded between zero and one. The closer the value to zero, the higher the indication that the cluster assignments are largely independent, while NMI close to one shows substantial agreement between the clusters. An NMI value of zero simply means that the label assignment is purely random. Figure \ref{figure:sfig1} shows the average NMI for different levels of noise and $d^2$, while Figure \ref{figure:sfig3} shows the distribution of NMI among 1000 runs.

It is evident from Figures \ref{figure:sfig1} and \ref{figure:sfig3} that \textit{MMCL} is able to achieve a high NMI value (about 0.8) when the $\epsilon$ is low. However, as expected, as the level of noise added to the response increases, it significantly affects the accuracy of clustering. With $\epsilon=4$, the clustering is almost done in a random fashion. We can also observe the effect of $d^2$. As mentioned earlier, smaller $d^2$ means that $\beta$s are closer and more similar to each other. Figures \ref{figure:sfig1} and \ref{figure:sfig3} confirm that as $d^2$ gets smaller, it is harder to correctly cluster the observations.

\begin{figure} [h]
	\begin{subfigure}{.5\textwidth}
		\centering
		\includegraphics[width=0.9\linewidth]{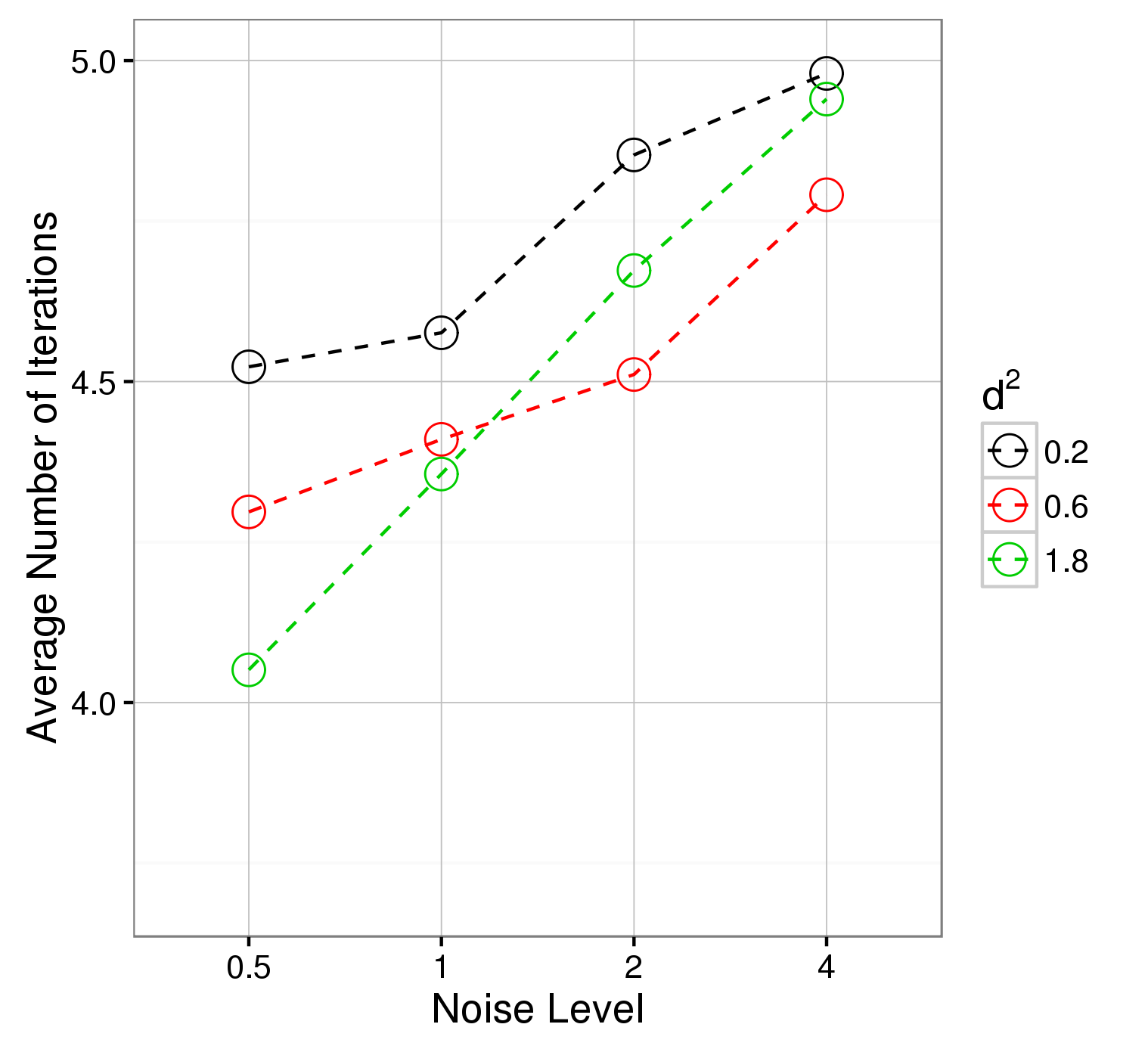}
		
		\caption{\footnotesize{}}
		\label{figure:MMCLe_sfig1}
	\end{subfigure}
	\begin{subfigure}{.5\textwidth}
		\centering
		\includegraphics[width=0.9\linewidth]{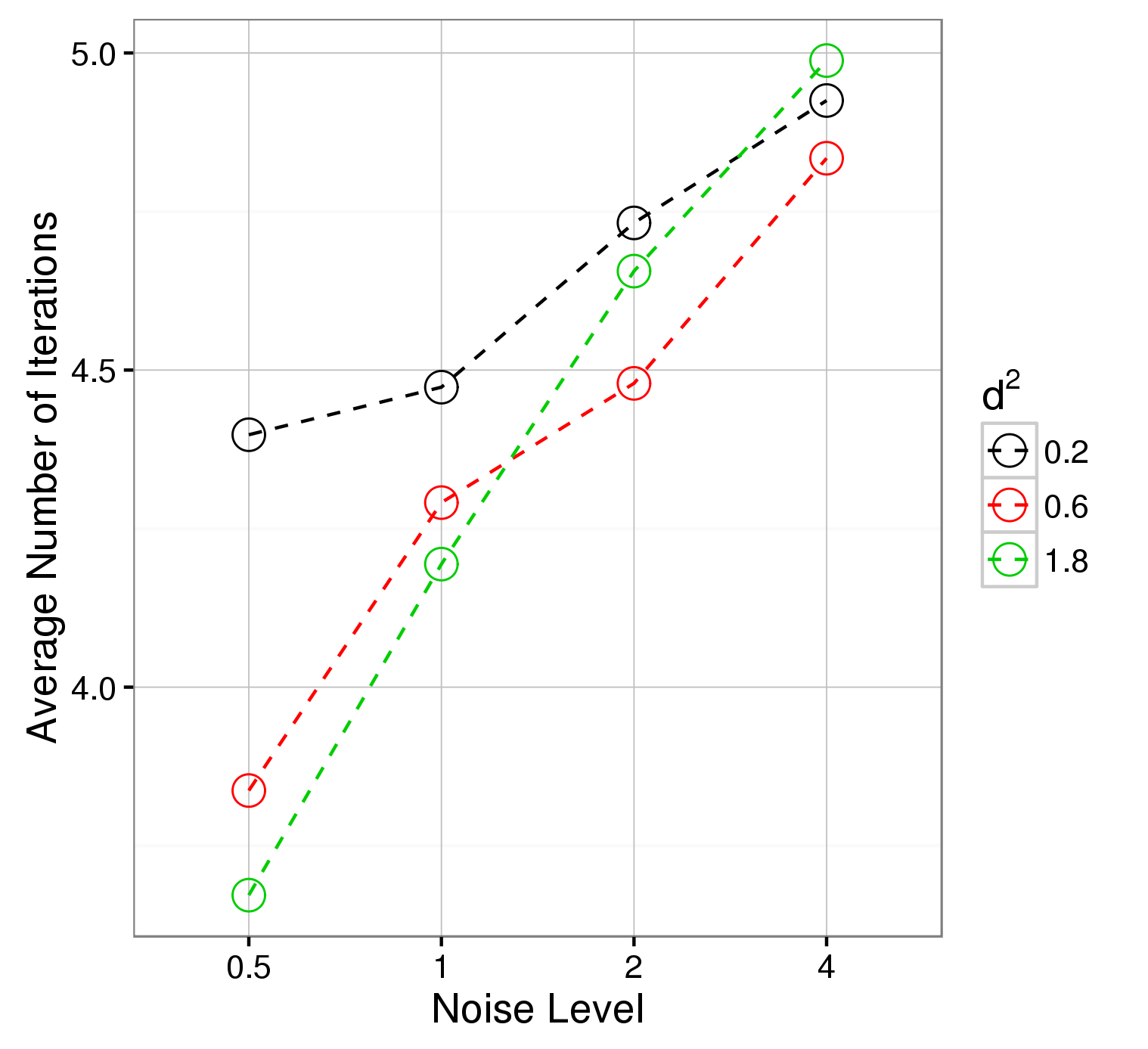}
		\caption{\footnotesize{}}
		\label{figure:MMCLe_sfig2}
	\end{subfigure}
	\caption{Evaluating \textit{MMCL} Efficiency in Iterations: (a)Average number of iterations for different noise levels and $d^2$ (random initial groups assignment). (b) Average number of iterations for different noise levels and $d^2$ using \textit{MMCL++}.}
\end{figure}

Figure \ref{figure:MMCLe_sfig1} shows the average number of iterations it took for \textit{MMCL} to converge under different scenarios. It can be seen that on average the algorithm converges within few iterations. Note that the number of iterations is highly affected by the stopping criteria. As mentioned earlier, the stopping criteria is based on the relative change in the overall AIC in two consecutive iterations, i.e. $|\frac{AIC-AIC_{Old}}{AIC_{Old}}|<\epsilon$. In this experiment, $\epsilon$ is set to 0.001 and the maximum number of iterations is set to 10.

To observe the impact of careful initial group selection (\textit{MMCL++}), we can compare Figure \ref{figure:sfig1} with \ref{figure:sfig2} and \ref{figure:sfig3} with \ref{figure:sfig4} and note that there is a slight improvement in the result when \textit{MMCL++} is utilized, especially when the noise level is low (0.5). Figures \ref{figure:MMCLe_sfig1} and \ref{figure:MMCLe_sfig2} compare the average number of iterations until convergence between random initial group selection (a) and \textit{MMCL++} (b). The graphs clearly show the effectiveness of \textit{MMCL++} in reducing the number of iterations, especially under low noise level. This reduction may not seem to be notable in this simulation experiment, but it could significantly reduce the run time when dealing with large datasets where every iteration may take considerable time.

\section{Deriving Recommendations under MMCL: Multi-objective Optimization}
\label{sec:MOO}

As noted earlier, the focus of this manuscript is to develop methods to improve the performance of individual stores by relying on a data-driven approach to internal benchmarking. In particular, the goal is to identify factors driving automotive dealership performance in comparison with ``similar" dealerships and relying on optimization to derive tailored recommendations. However, as noted by \citep{thomas1998process}, more than one performance outcome usually needs to be considered because stores are responsible for multiple and sometimes conflicting performance measures (e.g., sales and profits). In addition, it is often the case that KPIs are competing for resources and cannot be adjusted independently at will (e.g., cash flow constraints might force a dealer to choose between adding more new vehicle sales staff or more service technicians but not both). 

In the context of \textit{MMCL}, each cluster yields specific component models depending on certain KPIs (predictors) for each of dependent variables (e.g., one model for sales and another model for profit). Performance management entails deriving recommendations (i.e., suggested levels for the predictor variables) consistent with the dynamics identified by the component models to attain balanced performance. Conflicts in performance measures will be revealed in the form of common predictor variables for multiple dependent variables and opposing coefficients (e.g., raising inventory at a store can increase sales but also increases inventory costs and might compromise profitability). The question becomes, "how to adjust the KPIs in order to jointly improve the performance of the dependent variables"? Without loss of generality, we discuss methods for addressing this problem under the assumption that there are two performance measures of interest $y_1$ and $y_2$ (e.g., vehicle sales and profit). Essentially, we need an optimization formulation to optimize both $y_1$ and $y_2$. This is in fact a Multi-objective Optimization (MOO) problem with the possibility of multicollinearity (i.e., relations exist between KPIs or predictor variables) as well as potential relationships among dependent variables.

Let us suppose that we have a dataset with $n$ observations and two dependent variables: $y_1$ and $y_2$, with a set of KPI predictors ${\{x_1,x_2,\dots,x_p\}}, x_i \in {\rm I\!R}^n$. Let $X$ denote the matrix of covariates (the design matrix), $X \in {\rm I\!R}^{n.p}$, where there are relations between $x_i$s and ($y_1, y_2$). Moreover, we have to check to see if there are relations between the predictors ($x_i$s) themselves (i.e., multicollinearity), and consider (i.e., satisfy) these relations when formulating the MOO problem. This is the key requirement in formulating this problem. The reason is that without considering the multicollinearity among the independent variables, the model will not take any interdependencies into consideration and therefore it might produce ineffective results/recommendations.

\subsection{Formulating MOO}
The goal here is to find the optimal values for the actionable predictor KPI variables in order to jointly improve dependent variables $y_1$ and $y_2$. Let $x_i  \in {\rm I\!R}^n, i \in {\{1,2,...,p\}}$ denote a predictor KPI variable.
We can setup the multi-objective optimization problem as follows:
\begin{enumerate}[1)]
\item \textit{Regression: Model relationships between independent (i.e., KPIs) and dependent variables}\\
Regress $y_1$ and $y_2$ on $x_1, x_2, ...,x_p$ to obtain $f_{y_1}(x_1, x_2, \dots,x_p)$ and $f_{y_2}(x_1, x_2, ...,x_p)$. If linear regression is employed for modeling, the result is:
\begin{align}
y_1=f_{y_1}(x_1, x_2, ...,x_p)=\beta_0+\beta_1 x_1 +\beta_2 x_2 + \dots + \beta_p x_p + \epsilon_{y_1}\\
y_2=f_{y_2}(x_1, x_2, \dots,x_p)=\gamma_0+\gamma_1 x_1 +\gamma_2 x_2 + \dots + \gamma_p x_p + \epsilon_{y_2}
\end{align}

\item \textit{Multicollinearity: Regress each independent variable as a function of remaining independent variables }\\
Again, in the case of linear regression, we will have:
\begin{align}
x_i=f(x_1,x_2,\dots,x_r,y_1,y_2)=\alpha_{i0}+\alpha_{i1} x_1 +\alpha_{i2} x_2 + \dots + \alpha_{ip} x_p + \epsilon_{x_i} \quad \alpha_{ii}=0, \quad i \in {\{1,2,\dots,p\}}
\end{align}

\item \textit{Optimization: Formulation}\\
The MOO problem can be tackled using several different approaches. The classical means of solving such problems were primarily focused on scalarizing multiple objectives into a single objective \citep{deb2014evolutionary}. In many cases, there does not exist a single solution that simultaneously optimizes each objective. In that case, the objective functions are said to be conflicting, and there exists a (possibly infinite) number of Pareto optimal solutions. A solution is called nondominated, Pareto optimal, Pareto efficient or noninferior, if none of the objective functions can be improved in value without degrading some of the other objective values. There exist different solution philosophies and goals when setting and solving multi-objective optimization problems. Both exact and heuristic methods (e.g., genetic algorithms) have been extensively investigated in the literature. See \citep{deb2001multi} for a good overview on the topic.

Given that our independent and dependent variables are (mostly) continuous and we have two dependent variables ($y_1$ and $y_2$), there exists a Pareto curve of optimal solutions. Let us assume that the primary interest is to maximize $y_1$, while satisfying an acceptable value for $y_2$.

If we define 
% *************IMPORTANT CHANGE: Carefully Review*************
%$t = y_1=f_{y_1}(x_1, x_2, \dots,x_p)=\beta_0+\beta_1 x_1 +\beta_2 x_2 + \dots + \beta_p x_p + \epsilon_{y_1}$
$t =\beta_0+\beta_1 x_1 +\beta_2 x_2 + \dots + \beta_p x_p$, then we arrive at the following formulation:
\begin{subequations}
\begin{alignat}{2}
\max  \quad & t    \\
\text{s.t.} \quad & y_2 \geq \tilde{y_2}\\
		 & \left|t-(\beta_0+\beta_1 x_1 +\beta_2 x_2 + \dots + \beta_p x_p)\right| \leq{k\hat{\sigma}_{\epsilon_{y_1}}} \\
         &\left |y_2 -(\gamma_0+\gamma_1 x_1 +\gamma_2 x_2 + \dots + \gamma_p x_p)\right| \leq{k\hat{\sigma}_{\epsilon_{y_2}}} \\
		&\left |x_i -(\alpha_{i0}+\alpha_{i1} x_1 +\alpha_{i2} x_2 + \dots + \alpha_{ip} x_p)\right| \leq{k\hat{\sigma}_{\epsilon_{x_i}}} \quad \alpha_{ii}=0 \quad \forall i\\
%		& x_i \in {\{a\%-b\% \; percentile \; of \; x_i} \} \quad \forall i          
		& x_i \in {[x_i^{LowerBound}; x_i^{UpperBound}]} \quad \forall i          
\end{alignat}
\end{subequations}

The accuracy and effectiveness of the formulation results rely on the accuracy of the regression models.
Since the regression models cannot be assumed to be perfect, constraints (9b--9d)  are designed to take the imperfection of the regression models into account. For this purpose, we allow slack for the regression models derived constraints proportional to $k*\sigma$, where $\sigma$ denotes regression model standard error. Small values of $k$ lead to strict constraints, i.e., strong agreement with regression models at the risk of recommendations that limit performance. Constraints (9e) are optional and limit the decision variables to practical bounds. 
\end{enumerate}

\section{Case Study: Deriving Recommendations for Automotive Dealerships}
\label{sec:Case}

We focus here on applying the proposed \textit{MMCL} method for modeling the productivity of automotive dealerships across the U.S. for a particular OEM. The raw data mostly come from monthly financial statements of dealers to OEMs, along with dealership demographic information. The financial statements contain hundreds of input and output metrics regarding the various departments in the dealership (new vehicle sales, finance, parts/service, used department, body-shop etc), resources (staffing levels/tenure, service bays etc), inventory (new and used inventory, inventory mix, age of inventory etc), marketing (type/volume), and performance (sales, variable/fixed costs, and profits for each of the departments). To maintain confidentiality of information, we are not able to reveal the OMEM. There are 3074 dealerships with monthly data for five years (2010-1015) with 281 key performance indicators (KPIs) deemed important by the domain experts. The data is aggregated to construct a design matrix $X \in {\rm I\!R}^{(3074\times60)\times281}$. After cleaning the data and calculating the KPIs using the raw data, we found that about 6\% of the data is missing. We used a \textit{matrix completion via soft thresholding SVD} technique to impute the missing data, using the ``\textit{softImpute}" package in R  \citep{softImpute}. The variables are then standardized to carry a mean of zero and standard deviation of one for improved performance modeling.

\subsection{Applying MMCL to the Dealership Performance Problem}

Because of the large size of the dataset and specially the large number of predictors, Least Absolute Shrinkage and Selection Operator (LASSO) \citep{tibshirani1996regression} is used for regression modeling of both the sales as well as the profitability of each dealership, for each month. As discussed in Section \ref{sec:meth}, the main parameters should be selected before running the algorithm. In this case, the  parameters are: number of clusters $k$, selection of $k$ dealers for initializing the clusters, and LASSO regularizer ($\lambda$). To find the best values for these parameters, the data is split into training (first four years for each dealer), and testing (financial data from last year). The parameters are then selected using cross-validation, by evaluating the quality of the models on predicting the testing data. The parameters that produce the best result (highest $R^2$ value) on the testing data are selected. Once the parameters are selected, the \textit{MMCL} algorithm is applied to the dataset. It is observed that in most cases, the algorithm converges in less than 15 iterations.

\subsubsection{Results}
To evaluate the effectiveness of the algorithm, the data from each region is further divided into three groups based on the expected number of vehicle registrations in each dealer's territory. We present here results for one region of the U.S. Note that there are two dependent variables ($y_1$ denoting profitability or return on sales and $y_2$ denoting new vehicle sales effectiveness). Tables \ref{table:2} reports the effectiveness of the proposed \textit{MMCL} method versus employing a single LASSO regression model for each registration group.

\setlength{\tabcolsep}{4pt}
\begin{table}[h!]
	\centering
	\caption{Comparison of \textit{MMCL} vs. Single Linear Model Performance on Dealership Dataset}
	\begin{tabular}{ cc|c|c|c|c| } 
		\cline{3-6}
		& & \multicolumn{2}{c|}{Profitability} & \multicolumn{2}{c|}{Sales Effectiveness} \\ \hline
		\multicolumn{1}{ |c }{Group}  & \multicolumn{1}{ |c| }{Algorithm} & Model Parameters & $R^2$ & Model Parameters & $R^2$\\	\hline 
		\multicolumn{1}{ |c  }{\multirow{3}{*}{Low Registrations} } &
		\multicolumn{1}{ |c| }{\textit{MMCL}} & $k=2, \lambda = 0.04$ & 0.58 & $k=3, \lambda = 0.001$ & 0.28\\ [0.5ex]
		\multicolumn{1}{ |c  }{}                        &
		\multicolumn{1}{ |c| }{\textit{MMCL++}} & $k=2, \lambda = 0.04$ & \textbf{0.61} & $k=3, \lambda = 0.005$ & \textbf{0.28}\\ [0.5ex]
		\multicolumn{1}{ |c  }{}                        &
		\multicolumn{1}{ |c| }{Single Linear Model} & $\lambda$ = 0.0036 & 0.56 & $\lambda$ = 0.0029 & 0.14 \\ \hline 
		\multicolumn{1}{ |c  }{\multirow{3}{*}{Medium Registrations} } &
		\multicolumn{1}{ |c| }{\textit{MMCL}} & $k=2, \lambda = 0.02$ & 0.69 & $k=3, \lambda = 0.003$ & 0.19\\ [0.5ex]
		\multicolumn{1}{ |c  }{}                        &
		\multicolumn{1}{ |c| }{\textit{MMCL++}} & $k=2, \lambda = 0.02$ & \textbf{0.70} & $k=3, \lambda = 0.03$ & \textbf{0.20}\\ [0.5ex]
		\multicolumn{1}{ |c  }{}                        &
		\multicolumn{1}{ |c| }{Single Linear Model} & $\lambda$ = 0.032 & 0.57 & $\lambda$ = 0.0032 & 0.09\\ \hline
		\multicolumn{1}{ |c  }{\multirow{3}{*}{High Registrations} } &
		\multicolumn{1}{ |c| }{\textit{MMCL}} & $k=2, \lambda = 0.01$ & 0.54 & $k=2, \lambda = 0.006$ & 0.23\\ [0.5ex]
		\multicolumn{1}{ |c  }{}                        &
		\multicolumn{1}{ |c| }{\textit{MMCL++}} & $k=2, \lambda = 0.04$ & \textbf{0.54} & $k=2, \lambda = 0.03$ & \textbf{0.26}\\ [0.5ex]
		\multicolumn{1}{ |c  }{}                        &
		\multicolumn{1}{ |c| }{Single Linear Model} & $\lambda$ = 0.01 & 0.53 & $\lambda$ = 0.0017 & 0.11\\ 
		\hline
	\end{tabular}
	\label{table:2}
\end{table}

As the results suggest, especially in the case of $SE$, the \textit{MMCL} has improved the accuracy of the models in predictin
g both profitability and sales effectiveness. For example, for medium registrations group, using \textit{MMCL++}, we are able to achieve 26\% $R^2$ on the testing dataset for sales effectiveness in comparison with just 11\% for the case of a single model. This result also suggests that there is heterogeneity among dealers and by clustering them, one can improve the analysis and generate better recommendations to dealers for improving their performance. 

\subsection{Applying MOO for Dealership Performance Improvement}
\label{ch:proposal}
As noted earlier, there are two dealer performance characteristics of interest: Profitability ($P$: $y_1$) and Sales Effectiveness ($SE$: $y_2$). It is desired that both $y_1$ and $y_2$ be maximized for every dealer to improve the profitability of the dealership and satisfy the needs of the OEM in selling more new vehicles. 

To apply the MOO, we first regress $P$ ($y_1$) against $X$ to obtain:
\begin{equation}
P=f_{P}(x_1,\dots,x_p)= \beta_0+\beta_1 x_1 +\beta_2 x_2 + \dots + \beta_p x_p + \epsilon_{P}
\end{equation}
\noindent
where $f_{P}(x_1,\dots,x_p)$ is the objective function that we want to maximize (ignoring the error term $\epsilon_{P}$).
We then regress $y_2$ against $X$ to obtain the constraint that explains the relation between $y_2$ and $x_i's$:
\begin{equation}
SE=f_{SE}(x_1, x_2, \dots,x_p)=\gamma_0+\gamma_1 x_1 +\gamma_2 x_2 + \dots + \gamma_p x_p + \epsilon_{SE}
\end{equation}
\noindent
Also, to guarantee that $SE$ stays in the accepted range enforced by OEM, we add the following constraint:
\begin{gather*}
SE \geq \tilde{SE}
\end{gather*}
Lastly, we regress each $x_i$ against other $x_i$s, to account for the multicollinearity constraints. As explained in Section \ref{sec:MOO}, to consider the fact that the regression models are imperfect, we allow a slack of $k*\sigma_\epsilon$ for each regression model. The client also provided valid bounds for each of the predictor variables but are not reported here for confidentiality. Assuming $t = f_{P}(x_1,\dots,x_p)$, the final formulation takes the following form:
\begin{subequations}\label{eq:moo.dlr}
\begin{alignat}{2}
\max  \quad & t    \\
\text{s.t.} \quad & SE \geq \tilde{SE}\\
&   \left|P-(\beta0+\beta_1 x_1 +\beta_2 x_2 + \dots + \beta_p x_p)\right| \leq{k\hat{\sigma}_{\epsilon_{P}}} \\
         &\left |SE -(\gamma_0+\gamma_1 x_1 +\gamma_2 x_2 + \dots + \gamma_p x_p)\right| \leq{k\hat{\sigma}_{\epsilon_{SE}}} \\
&\left |x_i -(\alpha_{i0}+\alpha_{i1} x_1 +\alpha_{i2} x_2 + \dots + \alpha_{ip} x_p)\right| \leq{k\hat{\sigma}_{\epsilon_{x_i}}} \quad \alpha_{ii}=0 \quad \forall i        
\end{alignat}
\end{subequations}

\begin{figure}[H]
\centering
\begin{subfigure}{.4\textwidth}
\centering
\includegraphics[width=0.9\linewidth]{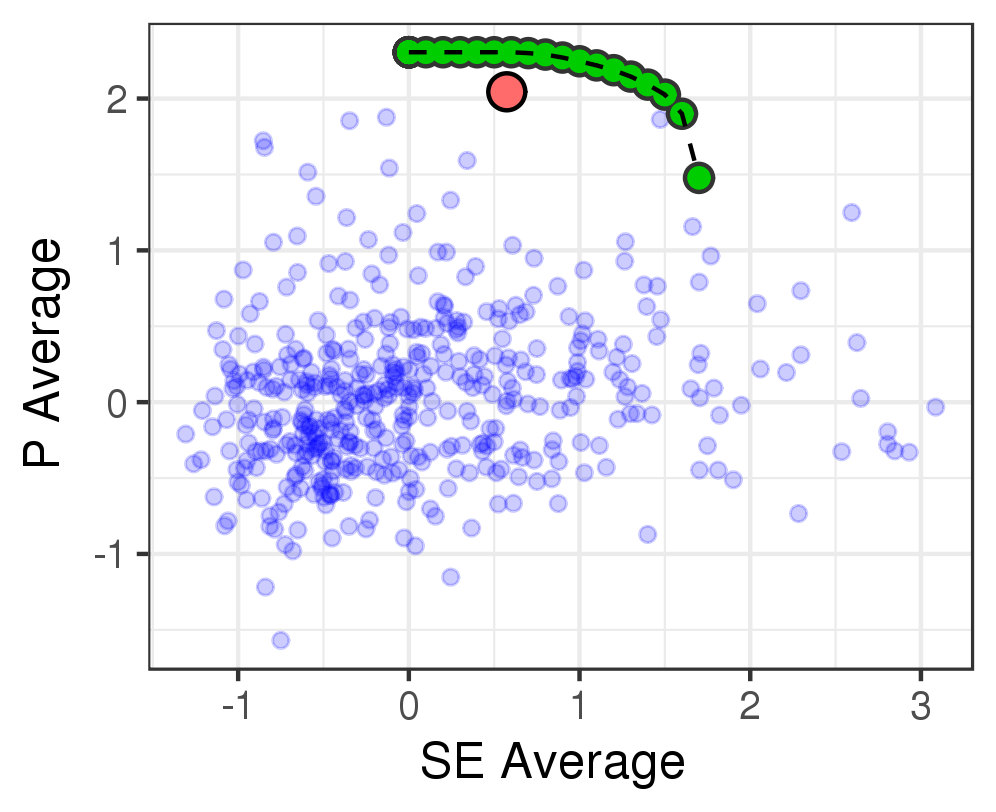}
\caption{\footnotesize{}}
\label{figure:pareto1}
\end{subfigure}%
\begin{subfigure}{.4\textwidth}
\centering
\includegraphics[width=0.9\linewidth]{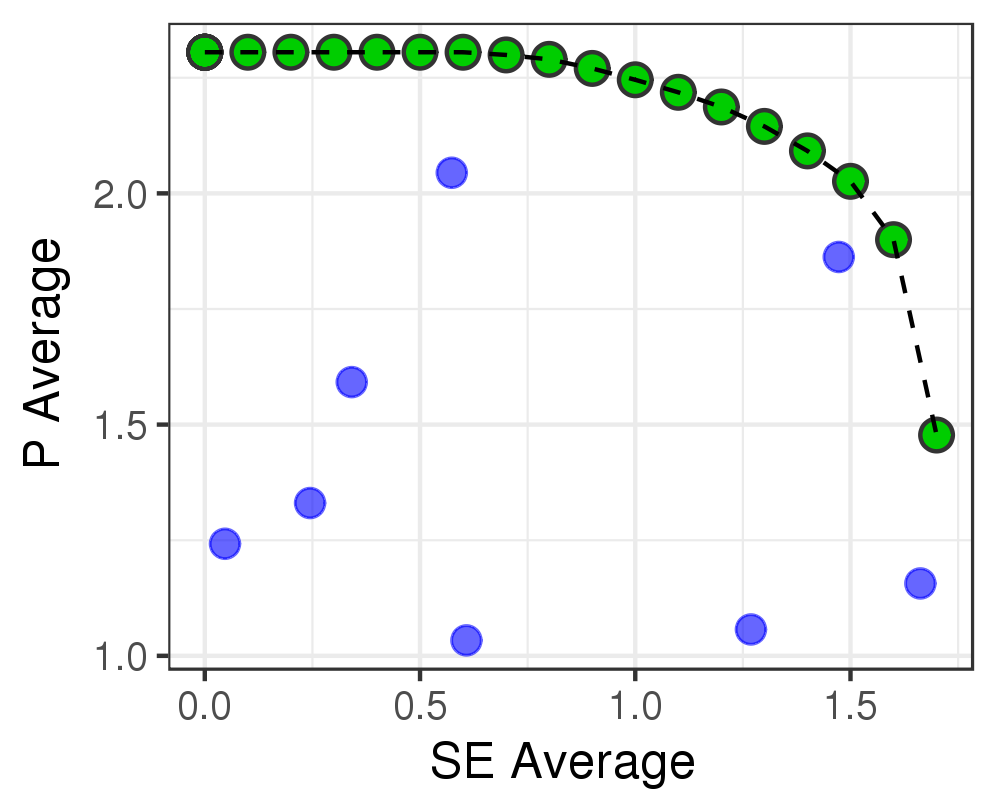}
\caption{\footnotesize{}}
\label{figure:pareto2}
\end{subfigure}
\caption{Pareto optimal frontier for a specific dealership group: (a) Blue dots report average $SE$ and $P$ for dealers; benchmark dealer is shown in red. (b) Zoomed-in region of Pareto optimal frontier}
\label{figure:pareto}
\end{figure}

\subsection{Generating Pareto Optimal Frontier}
To construct the Pareto optimal front for each dealership group, the formulation above is solved repeatedly by changing the value of $\tilde{SE}$ to obtain the Pareto optimal points for $P$ and $SE$, as reported in Figure \ref{figure:pareto1}. This result identifies what is potentially possible in terms of performance for the dealers within the reference group. As expected, the frontier also reveals the trade-off between how much profit a dealer can generate (using all the potential resources such as new and used vehicle sales, service, body shop, parts, etc.) versus new vehicle sales (not as profitable these days with respect to other dealership operations such as service and used vehicle sales).

\begin{figure}[H]
	\begin{subfigure}{.5\textwidth}
		\centering
		\includegraphics[width=1\linewidth]{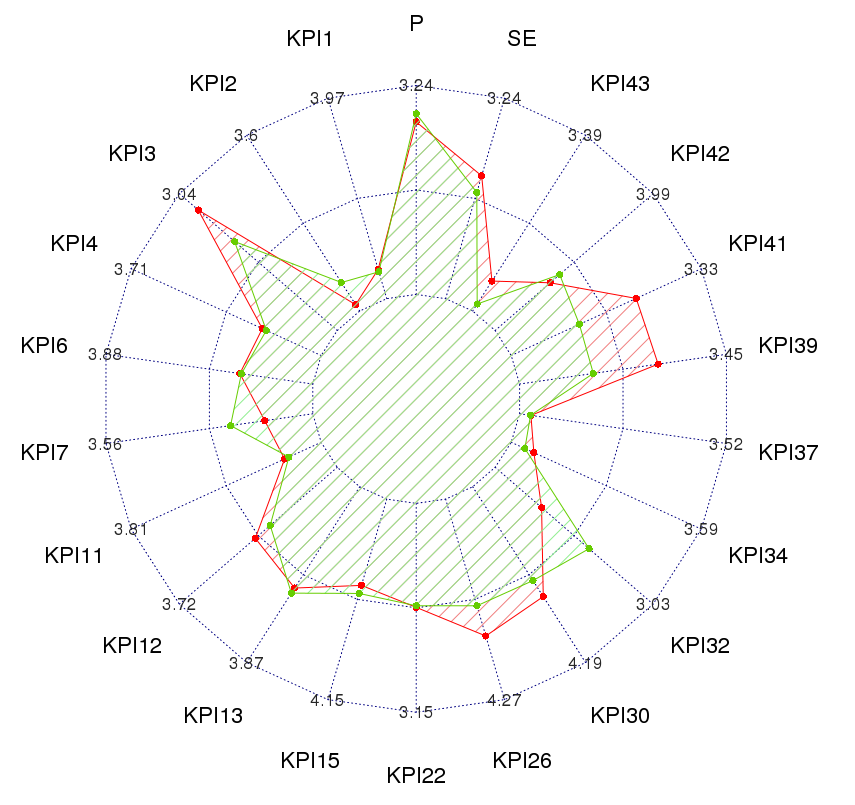}
		\caption{}
		\label{figure:radar.a}
	\end{subfigure}%
	\begin{subfigure}{.5\textwidth}
		\centering
		\includegraphics[width=1\linewidth]{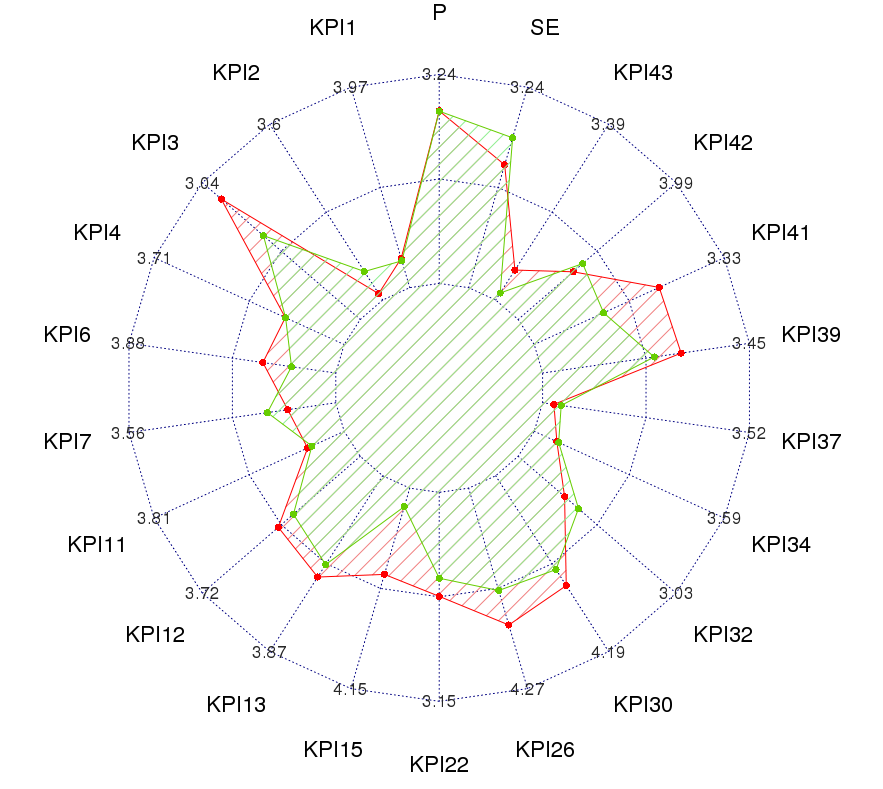}
		
		\caption{}
		\label{figure:radar.b}
	\end{subfigure}
	\captionsetup{justification=centering}
	\caption{Comparing derived recommendations for a reference dealership group (green) \\with the operations of a well performing dealer (red). (a) $\;\tilde{SE}=1.0$ (b) $\;\tilde{SE}=1.5$}
	\label{figure:radars}
\end{figure}

\subsection{Assessing the Quality of Recommendations Derived through MOO}
To further assess the effectives of the proposed methods, we compared the operational signature of a high performance dealership (shown in Figure \ref{figure:pareto1}) with the recommendations derived through MOO. Radar charts are generated for some of the important KPIs ($x_i$s) for different values of $\tilde{SE}$ (see Figure \ref{figure:radars}). The reasonable agreement between the two sets of KPI values between the high performance dealer and the derived recommendations further validate the effectiveness of our proposed algorithm. For example, it can be seen that if a dealer in the reference group wants to achieve a better $SE$ (on average), they have to on average lower KPI3, KPI26, KPI31, and KPI15, and increase KPI7 and KPI32. The recommended KPI levels can be compared with each dealer within the group in order to show the potential strengths and deficiencies of that dealer. It provides tailored guidance for the dealership management on how to manage and operate their business in order to simultaneously please the OEM by selling new vehicles (keep $SE$ in an acceptable range) and also increase their profits. 

\section{Conclusion}
\label{sec:conc}

Increasing availability of data combined with improvements in computational platforms and technology is enabling more comprehensive and in-depth data analysis in the world of business. In the retail sector, individual stores need to utilize the available data to improve both their efficiency and effectiveness for survival and dominance. 

Our objective was to develop a complete data-driven methodology for analyzing, processing, and modeling data from retail industry in order to understand the behavior of network of stores and provide scientific managerial guidance on how to improve and operate an individual as well as groups of stores. To achieve this goal, we addressed the problem of mixture models with group structure, and noted that this has not been addressed in the literature with the existence of a dependent variable (mixture of regressions with group structure). We proposed a solution to this problem called \textit{MMCL} for segmenting stores using model based clustering. \textit{MMCL} is an iterative, heuristic algorithm based on Competitive Learning to cluster groups of observations and provide a model for each group. It is a non parametric approach that can be combined with any underlying regression modeling technique. An extension to \textit{MMCL}, labeled \textit{MMCL++} is introduced to smartly select the initial groups when performing the clustering. \textit{MMCL++} increases both the accuracy and speed of the algorithm. The other contribution is the development of a multi-objective optimization (MOO) formulation to improve the profitability of the retail unit while controlling other performance metrics to meet the expectations of different stakeholders. Our MOO formulation is based on statistical relations between the decision variables and is designed to handle multicollinearity among the variables. The outcome of MOO provides recommended values for the decision variables (e.g. key store performance drivers) and can be used as managerial operational guidance for the stores.

The proposed methods are validated using synthetic experiments as well as data from a real-world automotive dealership network case study for a leading global automotive manufacturer. The results of both synthetic and real-world experiments proved the accuracy and effectiveness of our methodologies. We believe that this research is a good starting point for developing an intensive and complete process for benchmarking and managing the performance of retail stores.

There are several avenues for future research. We can extend the proposed methods to embrace a maximum likelihood approach to FMR with must-link constraints and also allow joint determination of the optimal number of clusters. Also, it can be combined with recommendation process, meaning to judge the models based on the quality of recommendations. This can be an online reinforcement learning framework that assesses the result of recommendation and utilizes the result and feedback to adjust and improve the models. As for MOO, the formulation can be extended to form a pure multi-objective optimization formulation that jointly optimizes both objective functions (rather than maximizing a single objective while imposing constraints on the performance of the remaining objectives). The proposed methods should also be further validated using data from other retail settings/applications.

\section*{References}
\small
\bibliographystyle{plainnat}
\bibliography{bibfile2}

\end{document}